\documentclass{article}

\usepackage{epsfig}
\usepackage{graphics}
\usepackage{graphicx}
\usepackage[centertags]{amsmath}
\usepackage{amsfonts}
\usepackage{amsthm}
\usepackage{amssymb}
\usepackage{url}
\usepackage{subfigure}

\begin{document}

\title{On Wave-Based Majority Gates \\ with Cellular Automata}

\author{Genaro J. Mart\'{\i}nez$^{1,2}$, Andrew Adamatzky$^2$ \\ Shigeru Ninagawa$^3$, Kenichi Morita$^4$}

\date{}

\maketitle

\begin{centering}
$^1$ Escuela Superior de C\'omputo, Instituto Polit\'ecnico Nacional, M\'exico. \\
\url{gjuarezm@ipn.mx} \\

$^2$ Unconventional Computing Lab, University of the West of England, Bristol, United Kingdom. \\
\url{andrew.adamatzky@uwe.ac.uk} \\

$^3$ Kanazawa Institute of Technology, Kanazawa, Japan. \\
\url{ninagawa@neptune.kanazawa-it.ac.jp} \\

$^3$ Hiroshima University, Hiroshima, Japan. \\
\url{km@hiroshima-u.ac.jp} \\
\end{centering}

\begin{abstract}
We demonstrate a discrete implementation of a wave-based {\sc majority} gate in a chaotic Life-like cellular automaton. The gate functions via controlling of patterns' propagation into stationary channels. The gate presented is realisable in many living and non-living substrates that show wave-like activity of its space-time dynamics or pattern propagation. In the gate a symmetric pattern represents a binary value 0 while a non-symmetric pattern represents a binary value 1. Origination of the patterns and their symmetry type are encoded by the particle reactions at the beginning of computation. The patterns propagate in channels of the gate and compete for the space at the intersection of the channels. We implement 3-inputs {\sc majority} gates using a {\sf W} topology showing additional implementations of 5-inputs {\sc majority} gates and one tree (cascade) {\sc majority} gate.


\end{abstract}

\begin{small}
\noindent Published in: {\sf WSPC Book Series in Unconventional Computing. Chapter 9: On Wave-Based Majority Gates with Cellular Automata. Handbook of Unconventional Computing, pp. 271-288 (2021). \url{https://www.worldscientific.com/doi/abs/10.1142/9789811235740_0009}}
\end{small}

\newpage
\section{Introduction}\label{sa_sec1}
Recent years have shown a growing interest to physical implementations of  {\sc majority} gates. There are three examples. First one is a spintronic device ---  {\em spin-wave majority gate} --- which is of a low space complexity and consumes ultra-lower power. Fischer et al.~\cite{fischer2017experimental} demonstrate a microwave device that can be constructed from majority gates using a trident topology. The device uses the interference of spin-waves, these waves are synchronized patterns of electron spin. The second example is the {\em plasmonic majority gate}, where propagate between a metal and a dielectric device. Dutta et al.~\cite{dutta2017proposal} implement plasmonic majority gates in a nanoscale cascadable photonic media. The information is encoded in the amplitude and phase of electric waves intensity.

Cellular automata computations they have been implemented frequently by atomic signals, inspired by von Neumann proof of universality of automaton with 29 cell-states \cite{burks1966theory}. Computation via reactions between particles --- gliders or mobile localizations --- was stimulated with the popular Conway's Game of Life where a lot of complex patterns emerge and interact \cite{rendell2016turing}. On the other hand, computation via competing patterns was proposed in \cite{martinez2010computation}.

Life-like family of rules are discrete analog of sub-excitable media and the Game of Life is not the unique rule with complex behaviour in this domain as was shown by Eppstein in \cite{eppstein2010growth}. There is a family of Life-like rules, where `cells never die' that means that the state `1' is an absorbing state. One of them is the family of {\it Life without Death} (LwD),\footnote{\url{https://www.conwaylife.com/wiki/OCA:Life_without_death}} studied by Griffeath and Moore  \cite{griffeath1996life}. In the LwD automaton we can observe propagation of patterns, formed due to rule-based restrictions on propagation similar to that in sub-excitable chemical media and slime mould {\it Physarym polycephalum} \cite{adamatzky2010physarum}. The LwD family of cell-state transition rules is an automaton equivalent of the precipitating chemical systems as was discussed in a phenomenological study of semi-totalistic and precipitating cellular automata \cite{adamatzky2006phenomenology}. In the precipitating cellular automata channels are constructed and activated by interactions of particles that imitate a propagation of thousands of live organisms across the channels, competing for the space. In \cite{martinez2010computation, martinez2008logical}, we have exploited topologies as {\sf T} or {\sf X} \cite{adamatzky2009hot} implemented to construct adders \cite{martinez2010majority}. To explore the potential of such devices, we drawn analogues with  spintronic and plasmonic devices with {\sc majority} gates \cite{dieny2020opportunities}. In present chapter we will adapt our previous designs with a {\sf W} topology.

Advantages to use {\sc majority} gates with respect to serial and traditional gates include the reduction of operations and space, increase the speed and perform parallel computation. Step by step some circuits designed with {\sc nand} and {\sc xor} gates are improved with {\sc majority} gates. For example, quantum dot cellular automata exploit frequently {\sc majority} gates to design circuits \cite{amlani1999digital, navi2010five, prakash2019new}. Previously we have implemented a half-adder and a binary adder based {\sc majority} gates with {\sf W} topology \cite{martinez2010majority}.

This chapter is organized as follows. Section 1 introduces the state of art of the importance of majority gates and cellular automata propagating information. Section 2 gives the base function (cellular automaton), description of the rule, global dynamics and statistical characteristics. Section 3 explains how the computation using {\sc majority} gates is implemented by competing patterns. Section 4 presents discussions and conclusions.

\section{Propagation patterns in Life-like rules}

Life-like rules domain displays a number of complex functions, some of them without travelling localizations. In \cite{martinez2010computation, martinez2008logical} we presented rules with chaotic behaviour but patterns playing a role of ``walls'' and stopping the `chaotic' universe's expansion. In this study, we focus  on the evolution rule $B2/S2345$, known also as $R(2,5,2,2)$ in cellular automata literature.

The evolution rule $B2/S2345$ is described as follows. Each cell $x \in \Sigma$ takes two states `0' (`dead') and `1' (`alive'), and updates its state depending on its eight $\cal V$ closest neighbours:

\begin{enumerate}
\item {\it Birth}: a central cell $x_{i,j}$ in state 0 at time step $t$ takes state 1 at time step $t+1$ if it has exactly two neighbours in state 1, $\Sigma_{i=0}^{{\cal V}-1} x = 2$.
\item {\it Survival}: a central cell $x_{i,j}$ in state 1 at time $t$ remains in the state 1 at time $t+1$ if it has two, three, four or five live neighbours, $\Sigma_{i=0}^{{\cal V}-1} x = 2 | 3 | 4 | 5$.
\item {\it Death}: all other local situations.
\end{enumerate}

Once a resting lattice is perturbed (few cells are assigned live states), patterns of states 1 emerge, grow and propagate on the lattice quickly. The rule $B2/S2345$ is classed as a chaotic function.

The global behaviour of rule $B2/S2345$ can be described by a mean field polynomial and its fixed points. Mean field theory is a proven technique for describing statistical properties of cellular automata without analyzing evolution spaces of individual rules. The method assumes that elements of the set of states are independent, uncorrelated between each other in the rule's evolution space. Therefore we can study probabilities of states in a neighbourhood in terms of probability of a single state (the state in which the neighbourhood evolves), thus probability of a neighbourhood is the product of the probabilities of each cell in the neighbourhood.

\begin{figure}
\begin{center}
\subfigure[]{\scalebox{0.6}{\includegraphics{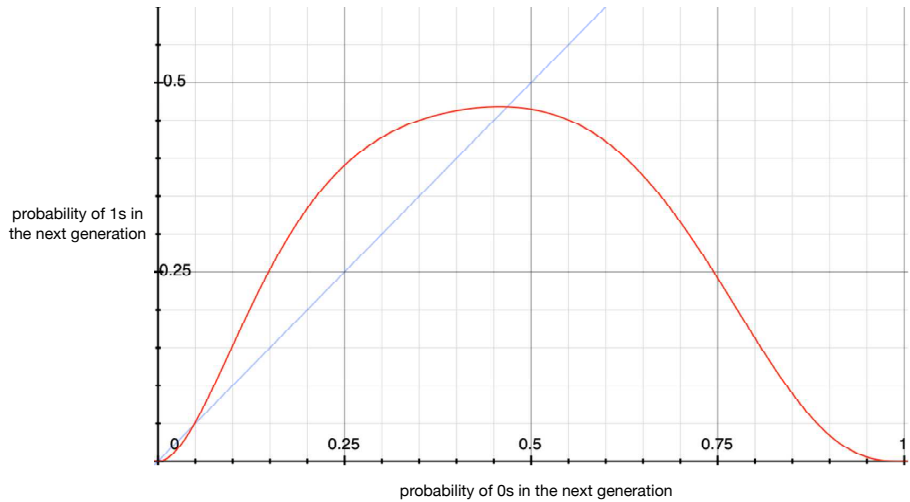}}} 
\subfigure[]{\scalebox{0.6}{\includegraphics{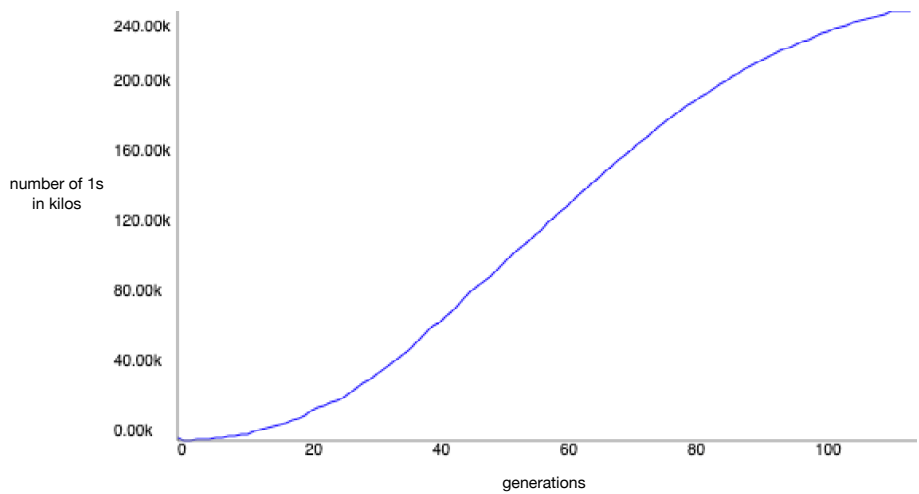}}}
\end{center}
\caption{(a) Shows the mean field approximation curve for the rule $B2/S2345$. Axis $x$ determines the probability to produce $0$s in the next generation and axis $y$ determines the probability to produce $1$s in the next generation. The rule has an unstable fixed point in 0.0476 and a stable fixed point in 0.468. (b) The plot displays the density of alive cells in a space of $700 \times 700$ cells starting from a random initial condition to 4\%, the density is fitted by a polynomial of order two.}
\label{meanField}
\end{figure}

McIntosh characterized a chaotic cellular automata with mean field approximation in \cite{mcintosh1990wolfram}. If the density plot crosses the diagonal and it has no tangencies then the function is chaotic. The mean field curve plotted in Fig.~\ref{meanField}a displays a stable fixed point in $f(x) = 0.0476$ that implies that a slow number of active cells will grow quickly, doubling the number of alive cells each two steps (Fig.~\ref{meanField}b), reaching the maximum concentration of alive cells  $f(x) = 0.4682$ where this maximum point is very close to the stable fixed point in $f(x) = 0.468$. It is quite interesting that this plot has similarities with the Game of Life polynomial where we have positive and negative tangencies related to complex behaviour of the rule.

\begin{figure}
\begin{center}
\subfigure[]{\scalebox{0.4}{\includegraphics{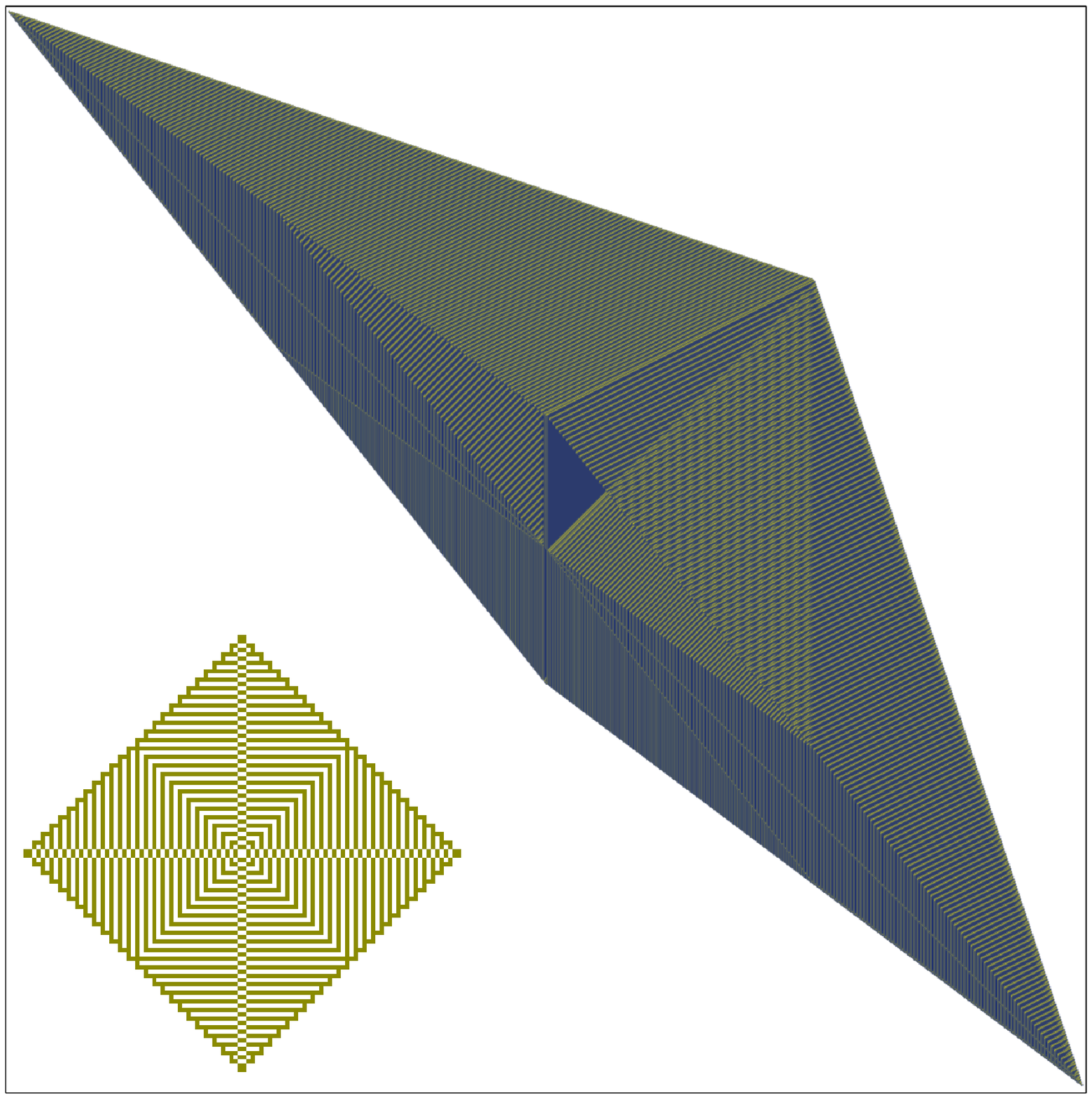}}} 
\subfigure[]{\scalebox{0.4}{\includegraphics{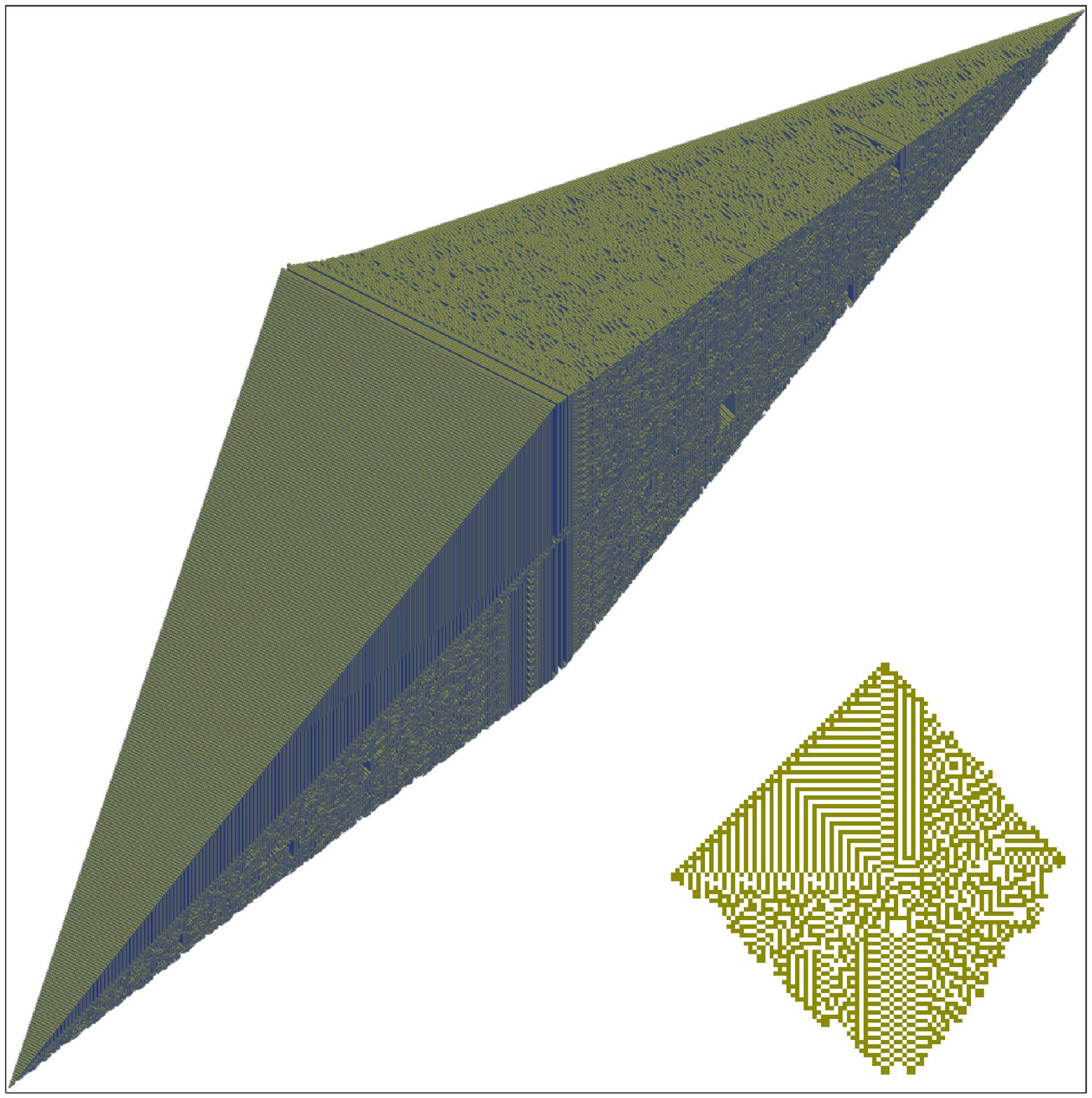}}}
\end{center}
\caption{Life-like rule $B2/S2345$ starting with a few cells in a 3D projection. (a) The evolution of a block of four alive cells. Evolution is periodic and symmetric expanding forever. The projection is on the $x$-axis after 320 steps. (b) {\sf L}-pentomino initial configuration displays an interesting global behaviour where chaos and periodic patterns coexist. The projection is on the $y$-axis after 300 steps.}
\label{fewcells}
\end{figure}

\begin{figure}
\begin{center}
\subfigure[]{\scalebox{0.36}{\includegraphics{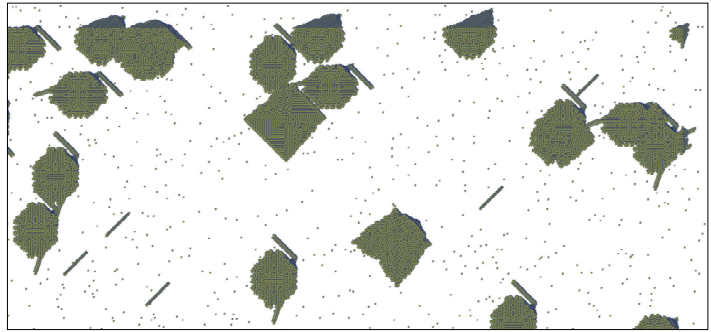}}} \hspace{0.3cm}
\subfigure[]{\scalebox{0.36}{\includegraphics{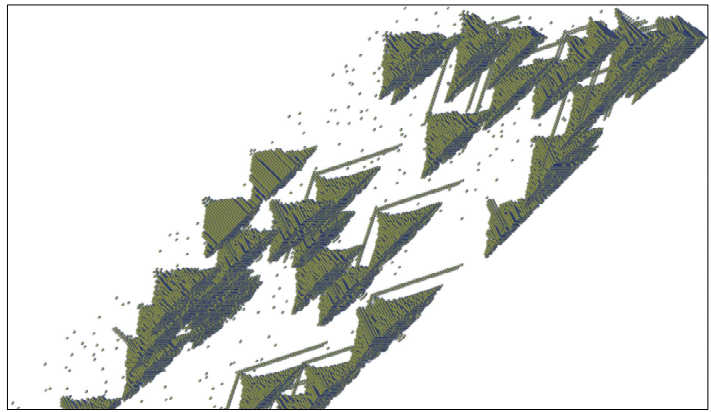}}} \hspace{0.3cm}
\subfigure[]{\scalebox{0.36}{\includegraphics{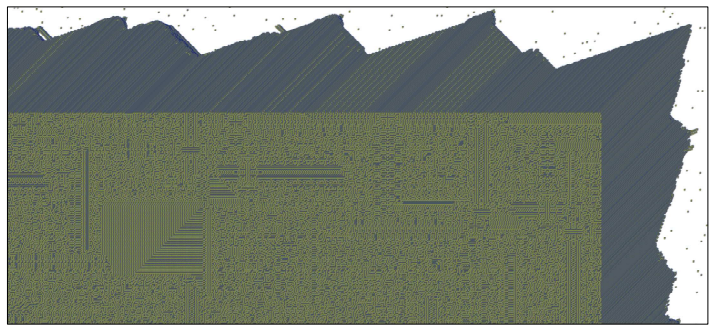}}}
\end{center}
\caption{Landscapes starting from random initial conditions to 4\% into a space of $700\times700$ cells. (a) After 30 steps we can see how nucleation of cells explodes but at the same time few mobile and stationary particles travel in the evolution space. (b) After 60 steps the chaos quickly expands along the $y$-axis. (c) The final global state where fixed configurations cover the whole evolution space.}
\label{random}
\end{figure}

The mean field polynomial for rule $B2/S2345$ is as follows: 

\begin{equation}
p_{t+1} = 14 p_t^2 q_t^3 (4 p_t^4 + 2 q_t^4 + 5 p_t^3 q_t + 2 p_t q_t^3 + 4 p_t^2 q_t^2).
\label{meanF}
\end{equation}

\noindent where $p_t$ is the probability of a cell being in state `1' at time step $t$ and its complement $q_t=(1-p_t)$ is the probability of the cell to be in state `0' at time step $t$.

Let us explore some specific initial conditions in details. Figure \ref{fewcells}a shows the evolution of the initial configuration $\left( \begin{array}{c} 11 \\ 11 \end{array} \right)$ after 320 generations with a population of 104,000 alive cells. The 3D projection is oriented along $x$-axis and thus the history of the evolution starts with an inclination of $45^{\circ}$. The base of this pyramid is the last step where the global behaviour shows a periodic pattern that expands symmetrically forever. While, Fig. \ref{fewcells}b displays the evolution of a {\sf L}-pentomino configuration $\left( \begin{array}{c} 1000 \\ 1111 \end{array} \right)$ later of 300 generations with a population of 95,048 alive cells. This evolution has a three-dimensional orientation on the $y$-axis. The evolution displays a non-trivial pattern where both emerge periodic and chaotic regions coexist fighting and expanding forever in the space, it is originated by the asymmetric formation of cells from the {\sf L}-pentomino. In this sense, the rule can be classified as a complex rule and not chaotic. At the same time, the evolution shows how some macro-cells emerge during the construction and they work as walls where a region has no communication with others. This feature is fundamental to construct channels of information to process binary data.

From random initial conditions we explore the universe with values close to the unstable fixed point. Fig \ref{random} shows the history when the automaton evolution starts with an initial condition at 4\% of active cells on a space of $700 \times 700$ squares. The first snapshot illustrates the first 30 steps where we can see how some few complex patterns emerge, such as still life, oscillators and particle patterns. As any chaotic function, the initial conditions are sensitive to small perturbations and can exploded into chaos. The second snapshot shows the history of 60 steps where particles travel at the speed of light $\frac{1}{c}$ and nucleations expand quickly and collide with other nucleations. For other densities of initial conditions it is impossible to observe such complex patterns. The third snapshot displays the final state of the global configuration where a mix of periodic and chaotic regions reach a stability. Therefore, this pattern is the result of collisions of particles and collisions of nucleations with a population of 264,578 alive cells after 190 steps.

\begin{figure}[th]
\begin{center}
\subfigure[]{\scalebox{0.95}{\includegraphics{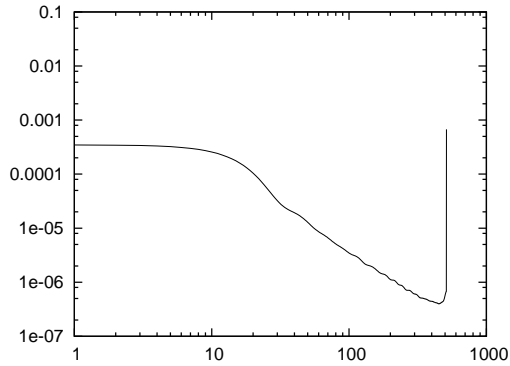}}} 
\subfigure[]{\scalebox{0.95}{\includegraphics{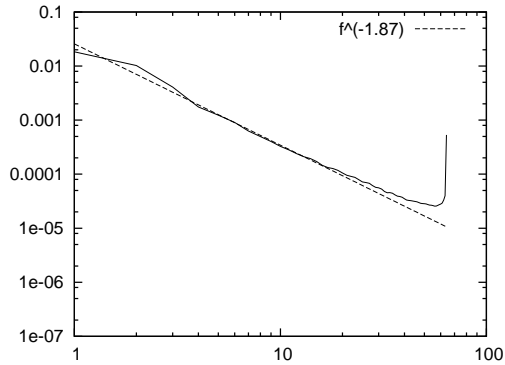}}}
\end{center}
\caption{Power spectra calculated from the evolution starting from random condition (a) at 1\% over 1,024 steps and (b) 128 steps. They are calculated in a square area of $50 \times 50$ cells among $700 \times 700$ cells. The $x$-axis is the frequency $f$, $y$-axis is power. The solid line in the right is the least square fitting of the data in the range of $f = 1 \sim 10$.}
\label{PowerSpec}
\end{figure}

The Game of Life shows $1/f$ noise in the evolution starting from random condition \cite{ninagawa1998flu}. On the other hand, $B2/S2345$ shows the Lorentzian spectrum (Fig.~\ref{PowerSpec}a) in which the power is flat at frequencies lower than some frequency determined by the time constant of the system. That implies that $B2/S2345$ has a finite relaxation time in its behaviour. This is caused by the characteristic behaviour of $B2/S2345$ that the patterns are fixed after the nucleations have finished.

The power spectra, however, differ in areas where the diffusive waves propagate in the channels of majority gate. The exponent of the power spectra is about -2, on condition that the observed steps are not long; e.g. 128 steps. This means that the behaviour of each cell is similar to Brownian motion. The power spectra with exponent close to -2 are observed also in the evolution from random configuration provided that the observed steps are not long (Fig.~\ref{PowerSpec}b). These results suggest that the behaviour of $B2/S2345$ is essentially similar to Brownian motion from the viewpoint of spectral analysis.

The most significant difference between $B2/S2345$ and Life is the existence or non-existence of `sheath' along the pathway of the signal. While in Life the signals are substantiated by bare propagating patterns and they move ahead in vacuous space, in $B2/S2345$ the pathway of the signal is covered with sheath constructed with stationary structures. Brownian-motion like power spectra are observed in another sheath-type CA, Langton's self-reproducing loop as well~\cite{ninagawa2021dynamics}. The existence or non-existence of sheath covering the pathway of signal might make a difference between Brownian motion and $1/f$ noise.

\section{{\sc Majority} gates by competing patterns}

{\sc Majority} gates by competing patterns in cellular automata were introduced in \cite{martinez2008logical, martinez2010majority, martinez2010computation} in {\sf X} topology.\footnote{\url{https://www.comunidad.escom.ipn.mx/genaro/Diffusion_Rule/Life_B2-S2345.html}} In this section, we will explore how {\sf W} topology will work. So, first we explore the universe of the Life-like rule $B2/S2345$.

\begin{figure}[th]
\begin{center}
\subfigure[]{\scalebox{0.38}{\includegraphics{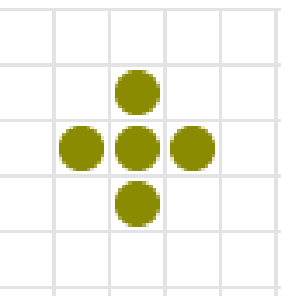}}} \hspace{0.8cm}
\subfigure[]{\scalebox{0.38}{\includegraphics{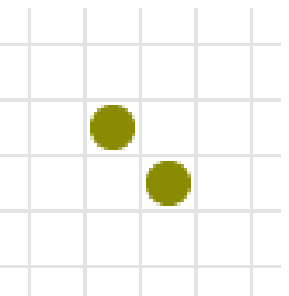}}} \hspace{0.8cm}
\subfigure[]{\scalebox{0.38}{\includegraphics{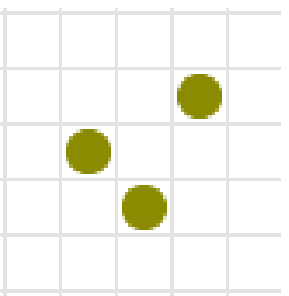}}} \hspace{0.8cm}
\subfigure[]{\scalebox{0.38}{\includegraphics{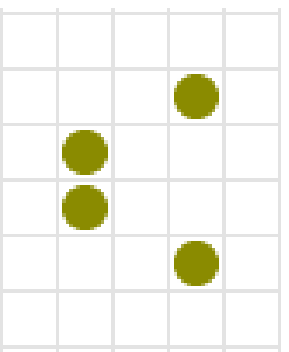}}}
\end{center}
\caption{Primitive patterns emerging in Life-like rule $B2/S2345$. (a) still life, (b) blinker, (c) oscillator, (d) particle.}
\label{primitivePatterns}
\end{figure}

Figure \ref{primitivePatterns} shows the very small universe of complex patterns that emerge in the Life-like rule $B2/S2345$. We saw in Fig.~\ref{fewcells} that small perturbations related to its unstable fixed point mean field polynomials induce chaotic behaviour. Four primitive complex patterns are one type of a still life, two types of oscillators and one particle/glider. The Tab.~\ref{tablePatterns} shows basic properties of these complex patterns.

\begin{table}[th]
\centering
\caption{Properties of primitive structures in Life-like rule $B2/S2345$.}
{\begin{tabular}{|c|c|c|c|c|c|}
\hline
structure & mass & volume & period & displacement & speed \\
\hline
still life & 5 & $3^2$ & 1 & 0 & 0 \\
blinker & 2 & $2^2$ & 2 & 0 & 0 \\
oscillator & 3 & $3^2$ & 2 & 0 & 0 \\
particle & 4 & $3 \times 4$ & 1 & 1 & $1/c$ (speed of light) \\
\hline
\end{tabular}}
\label{tablePatterns}
\end{table}

\begin{figure}[th]
\centering
\includegraphics[width=0.1\textwidth]{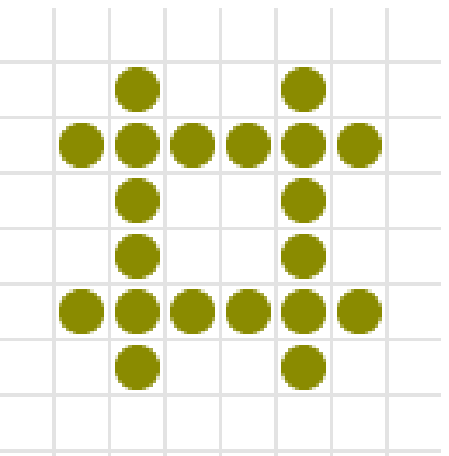}
\caption{Indestructible pattern in $B2/S2345$. It is a still life configuration composed by concatenation of four primitive still life patterns (Fig. \ref{primitivePatterns}a). }
\label{4StillLife}
\end{figure}

Particularly the rule $B2/S2345$ has an indestructible pattern composed by concatenation of four still life patterns forming a symmetric still life (Fig. \ref{4StillLife}). This pattern is very useful for our constructions because we can design wires where patterns shall propagate inside.

To construct a wire we need to define the minimum size where a colony of indestructible patterns does not form new active cells yet is conducive to propagation of particles. The particle has a size of four cells while the indestructible still life has a size of six cells, thus $6$ {\sf mod} $4 = 2$ yields the number of combinations where the particle can be encoded in a wire without exploding in chaos.

\begin{figure}[th]
\begin{center}
\subfigure[]{\scalebox{0.35}{\includegraphics{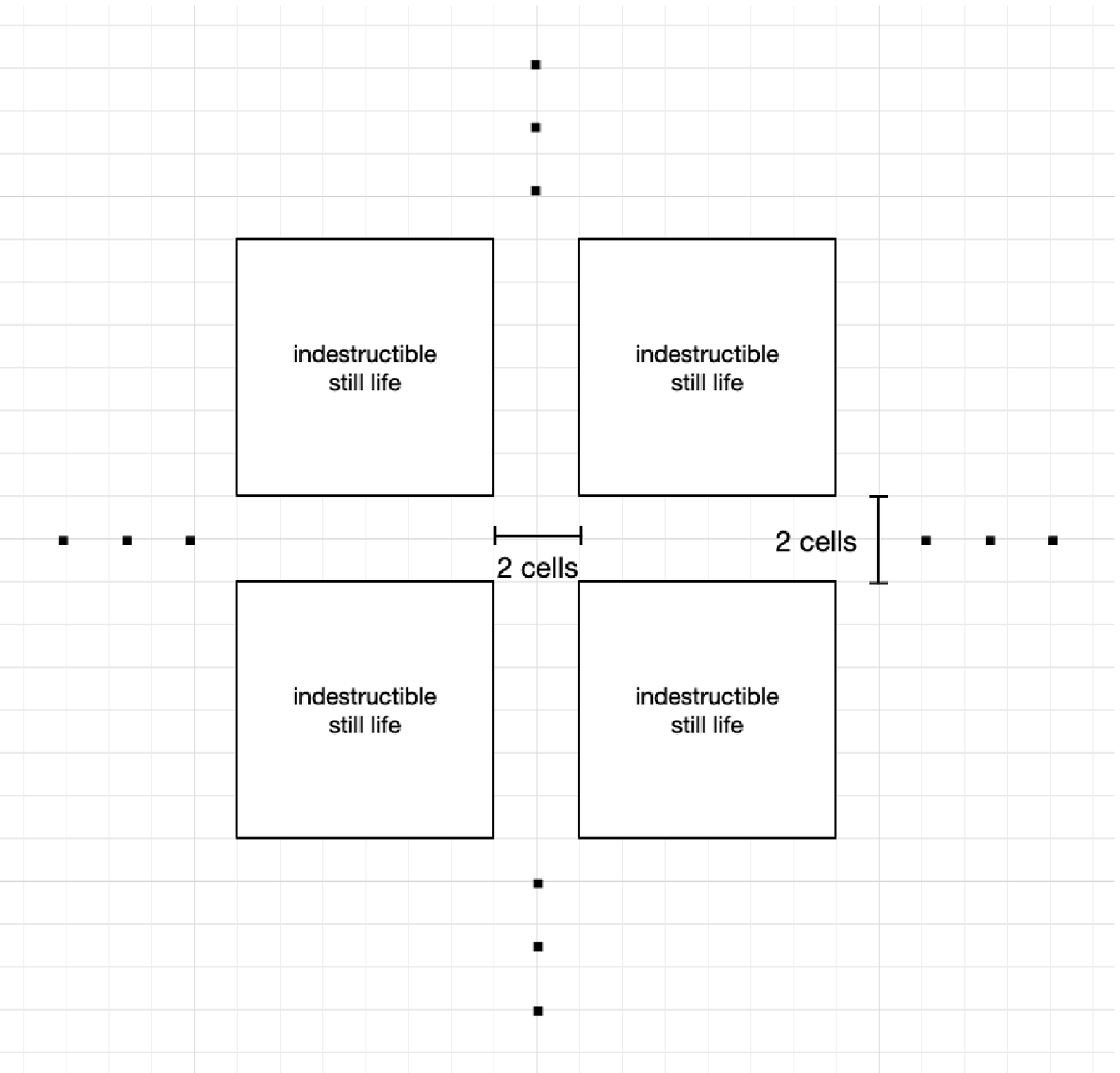}}} 
\subfigure[]{\scalebox{0.35}{\includegraphics{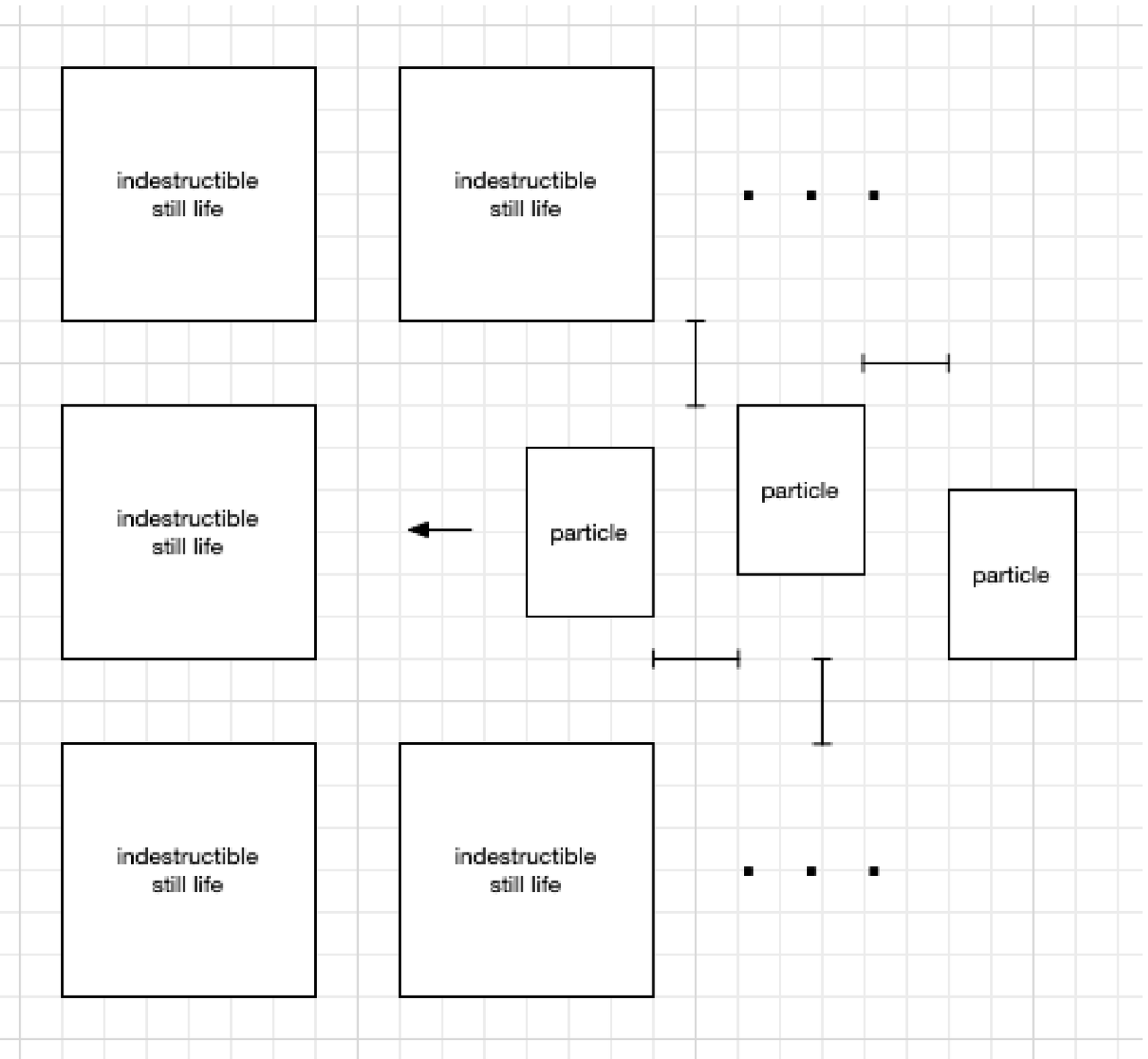}}}
\end{center}
\caption{(a) Minimal space to construct stable patterns formed by colonies of indestructible still life configurations. (b) The minimal space where a channel can be designed and particles can travel without perturbing the medium.}
\label{architectureWire}
\end{figure}

Figure \ref{architectureWire}a shows the minimum space between indestructible patterns to construct a colony of them (like an agar configuration) and they cannot be disturbed from any reaction. A particle preserves the same distance of two cells to travel freely without producing more information Fig. \ref{architectureWire}b. If two particles are traveling continually they shall preserve the distance of two cells to avoid an undesirable reaction. Finally, the wire where patterns will propagate is defined by a size of three indestructible still life configurations. 

The easiest way to control patterns propagating in a non-linear medium circuit is to constrain them geometrically. Constraining the media geometrically is a common technique used when designing computational schemes in spatially extended non-linear media. For example `strips' or `channels' are constructed within the medium (e.g. excitable medium) and connected together. Fronts of propagating phase (excitation) or diffusive waves represent signals, or values of logical variables. When fronts interact at the junctions some fronts annihilate or new fronts emerge. The propagation in the output channels represent results of the computation.

\begin{figure}[th]
\begin{center}
\subfigure[]{\scalebox{0.375}{\includegraphics{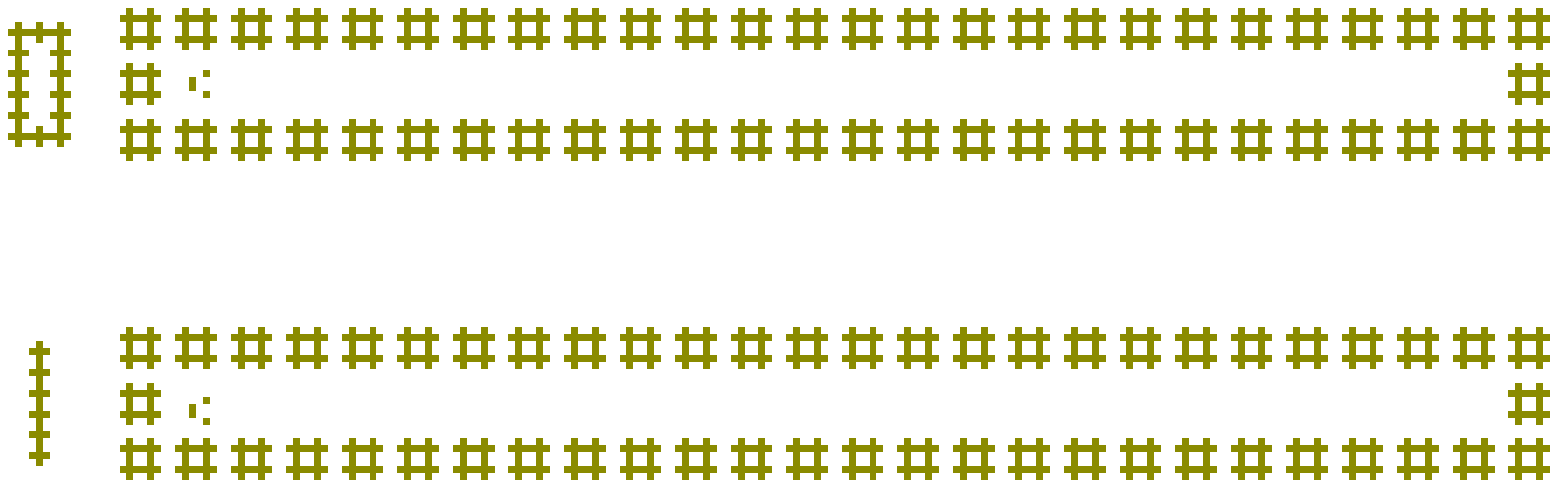}}} 
\subfigure[]{\scalebox{0.375}{\includegraphics{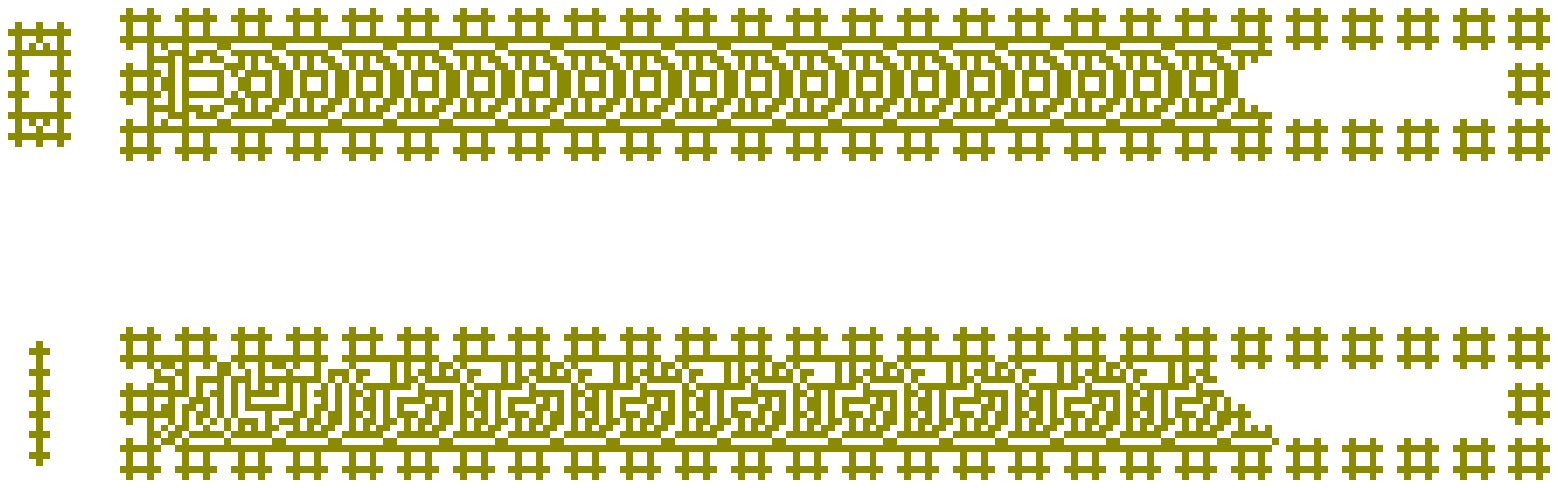}}}
\end{center}
\caption{Wires constructed with indestructible still life patterns in Life-like rule $B2/S2345$. (a) The basic structure of an empty channel with a particle initialized to encoded a symmetric (up) and asymmetric pattern (down). (b) The symmetric pattern propagating as a wave on the channel represents value 0 (up) and the asymmetric pattern represents value 1 late (down) of an excitation derived from a particle collision.}
\label{channel}
\end{figure}

Boolean values are represented by the position of particles, positioned initially in the middle of channel, value 0, or slightly offset, value 1 (Fig.~\ref{channel}a). The initial positions of the particles determine outcomes of their reaction. Particle, corresponding to the value 0 is transformed to a regular symmetric pattern, similar to frozen waves of excitation activity. Particle, representing signal value 1, is transformed to transversally asymmetric patterns (Fig.~\ref{channel}b). Both patterns propagate inside the channel with constant, advancing a unit of channel length per step of discrete time and  patterns repeat every 16 steps.

\begin{table}[th]
\centering
\caption{True table for a three-input {\sc majority} gate.}
{\begin{tabular}{|c|c|c|c|}
\hline
\multicolumn{4}{|c|}{3-input majority gate} \\
\hline
input & output & input & output \\
\hline
000 & 0 & 100 & 0 \\
001 & 0 & 101 & 1 \\
010 & 0 & 110 & 1 \\
011 & 1 & 111 & 1 \\
\hline
\end{tabular}}
\label{3i-majTable}
\end{table}

Minsky describes the {\sc majority} gate with disjunctive and conjunctive normal logical propositions for three inputs $A,B,C$ in \cite{minsky1967computation}, as follows:

\begin{equation}
{\sc MAJ}(A,B,C) = (A \wedge B) \vee (A \wedge C) \vee (B \wedge C)
\end{equation}

\noindent where the result is precisely the most frequent value on such variables. This way, we have that for three inputs $|\Sigma|^3$ is the number of outputs, as we can see in Table \ref{3i-majTable}.

\begin{figure}
\centering
\includegraphics[width=0.6\textwidth]{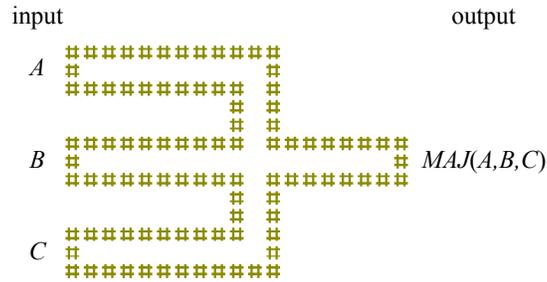}
\caption{A scheme to implement a {\sf W} architecture for 3-input {\sc majority} gate in the cellular automaton Life-like rule $B2/S2345$. The length of channels are defined internally (with still life patterns) by 72 cells for every input and the central channel that includes the output 138 cells, the size is defined by 10 cells. In average, after 185 generations the computation is done. The whole volume of this device is defined by a square of $150\times102$ cells.}
\label{WaveMajorityGateForm}
\end{figure}

\begin{figure}
\begin{center}
\subfigure[]{\scalebox{0.23}{\includegraphics{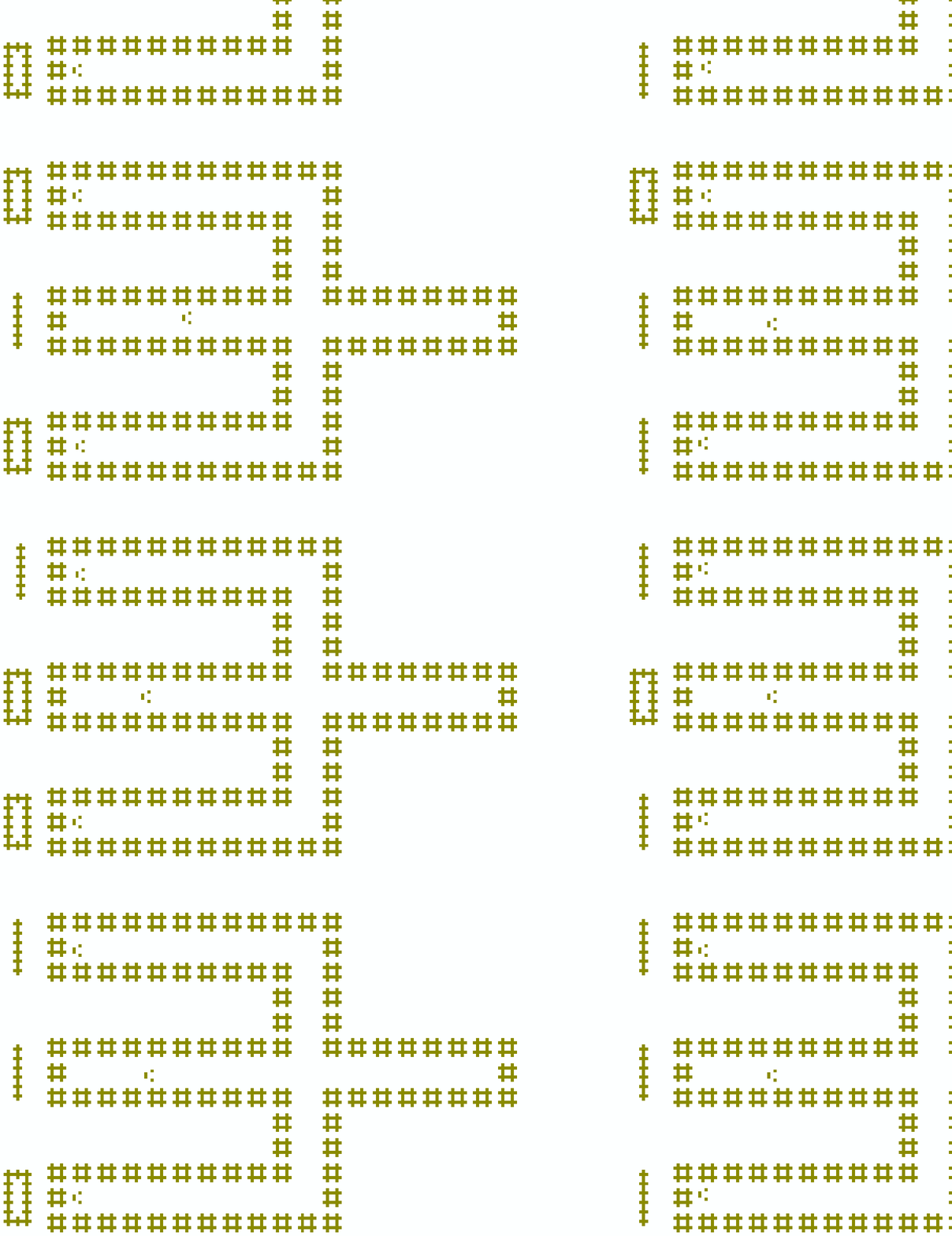}}} 
\subfigure[]{\scalebox{0.23}{\includegraphics{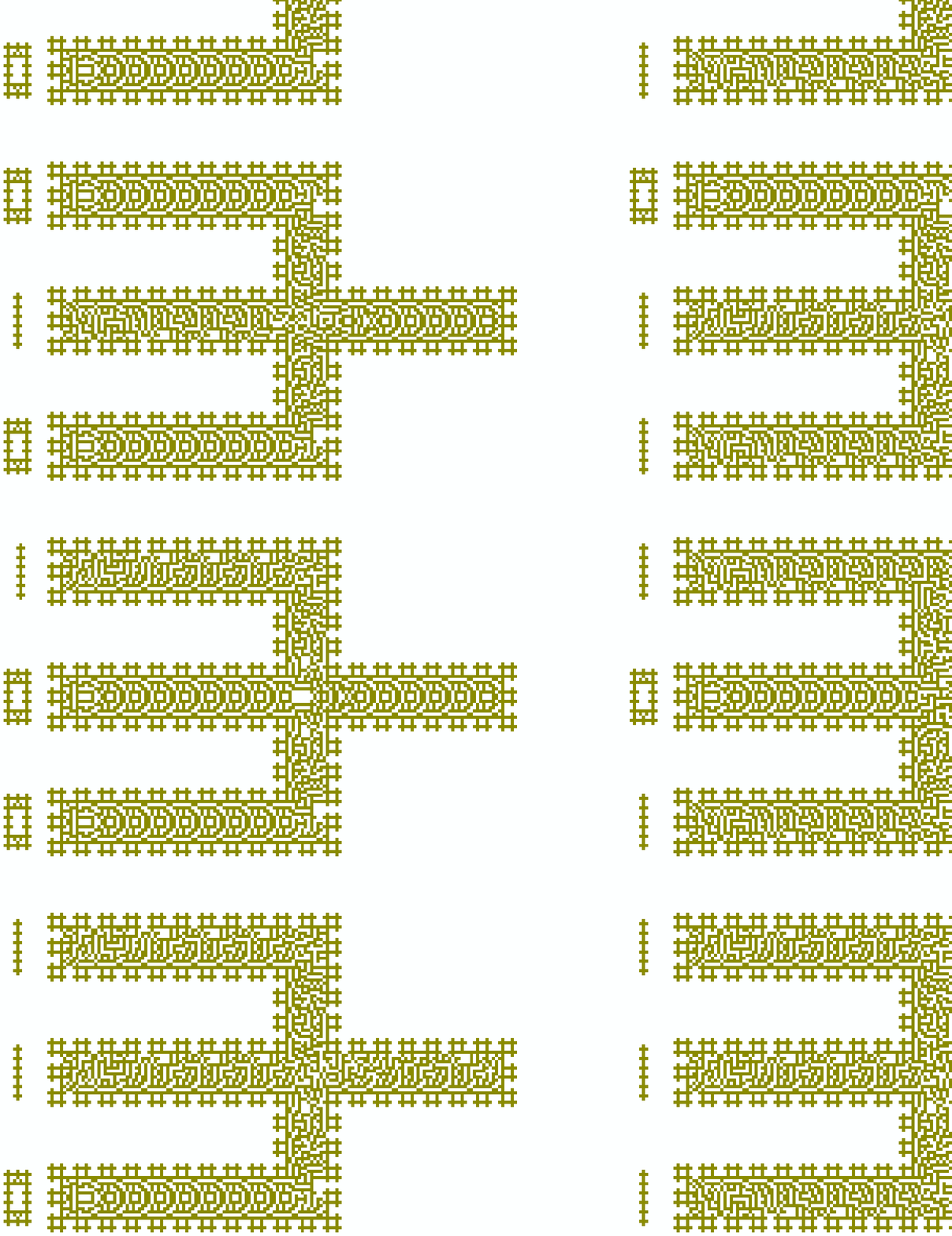}}}
\end{center}
\caption{Implementation of a 3-input {\sc majority} gate by competing patterns.}
\label{WaveMajorityGate}
\end{figure}

Conventional {\sc and, or, not} gates are easy to implement with competing patterns, see \cite{martinez2010computation}. However, when implementing non-serial logical gates it is more convenient to work with {\sc majority} gates, as they can be extrapolated to hot ice computers \cite{adamatzky2009hot} or slime mould in a more realistic way \cite{adamatzky2010physarum}, and also to molecular computing \cite{gao2017implementation}, spintronic technology \cite{fischer2017experimental, radu2015spintronic} or plasmonic devices \cite{dutta2017proposal}. Particularly, these kinds of gates are used to design a diversity of circuits based {\sc majority} gates in quantum dot cellular automata, see \cite{prakash2019new, navi2010five, amlani1999digital}). As has been demonstrated the use of {\sc majority} gates increase the speed and performance of computation in novel and updated algorithms, see \cite{pudi2017majority}.

A {\sc majority} gate in $B2/S2345$ in {\sf W} topology can be described as follows. The gate has three parallel orthogonal inputs: West channels, and one output: East channel. Three propagating patterns, which represent inputs, collide at the cross-junction of the gate. The resultant pattern is recorded at the output channel, as is illustrated in Fig.~\ref{WaveMajorityGateForm}.

Figure~\ref{WaveMajorityGate} shows the implementation of 3-input {\sc majority} gates with {\sf W} topology by competing patterns using the Life-like cellular automaton $B2/S2345$. The left side (Fig \ref{WaveMajorityGate}a) displays the initial condition where every particle defines binary values for each input and the right side (Fig \ref{WaveMajorityGate}b) displays the result after 178 generations.

One can also implement a 5-input {\sc majority} gate by competing patterns. This gate can be expressed with disjunctive and conjunctive normal logical propositions for five inputs $A,B,C,D,E$ as follows:

\begin{align}
{\sc MAJ}(A,B,C,D,E) &= (A \wedge B \wedge C) \vee (A \wedge B \wedge D) \vee (A \wedge B \wedge E) \vee \nonumber \\
&\qquad {} (A \wedge C \wedge D) \vee (A \wedge C \wedge E) \vee (A \wedge D \wedge E) \vee \nonumber \\
&\qquad {} (B \wedge C \wedge D) \vee (B \wedge C \wedge E) \vee (B \wedge D \wedge E) \vee \nonumber \\
&\qquad {} (C \wedge D \wedge E).
\end{align}

\begin{figure}
\centering
\includegraphics[width=1\textwidth]{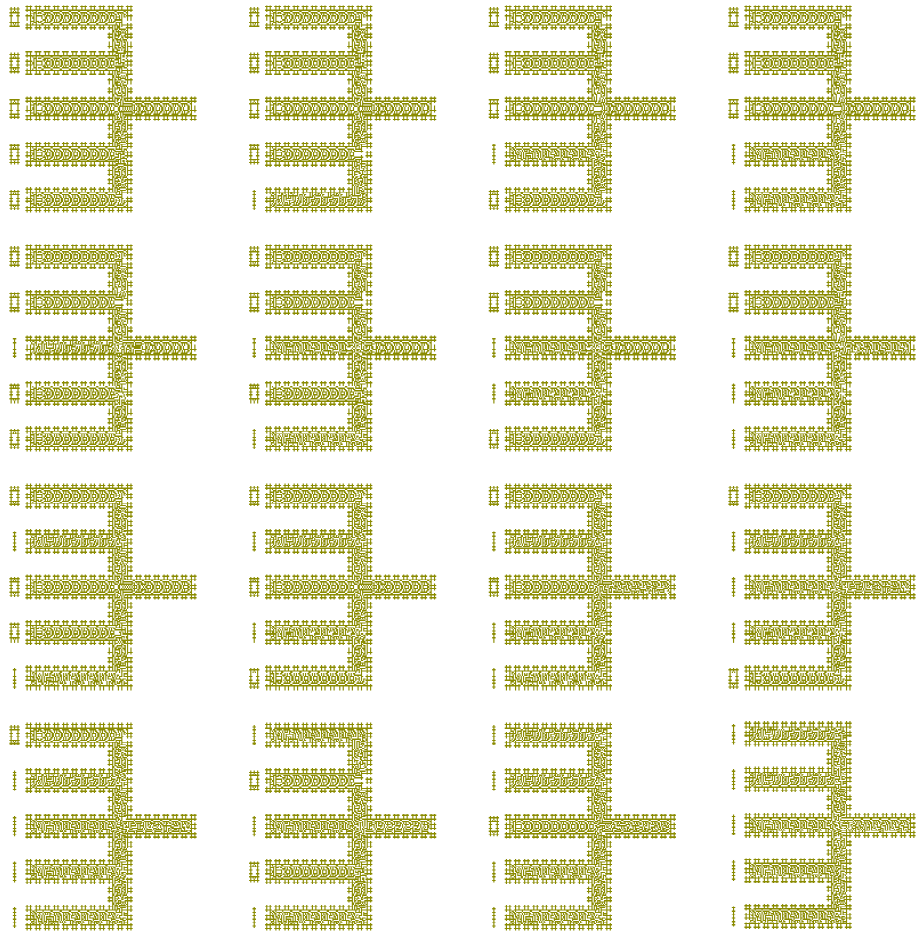}
\caption{Implementation of a 5-input {\sc majority} gate by competing patterns in $B2/S2345$.}
\label{5WaveMajorityGate}
\end{figure}

Figure \ref{5WaveMajorityGate} displays the implementation of 5-input {\sc majority} gate in $B2/S2345$. With this number of inputs we have $|\Sigma|^5$ operations. The evolutions are symmetric and therefore it is not necessary to calculate the whole set of gates. This way, the input $MAJ(0,1,0,0,0)$ is symmetric to its codification and representation to the input $MAJ(0,0,0,1,0)$ and so on. This 5-input {\sc majority} gate architecture is implemented in a volume of $150 \times 183$ cells and every gate needs five particles to start the reaction and the propagation of patterns.

Of course, increasing the number of $n$-inputs in {\sc majority} gates might lead to problems with synchronization of collisions. This is somewhat reflected in designs of quantum dot cellular automata where adders are constructed from  5-input {\sc majority} gates~\cite{sasamal2016optimal, wang2018novel}.

\begin{figure}
\centering
\includegraphics[width=1\textwidth]{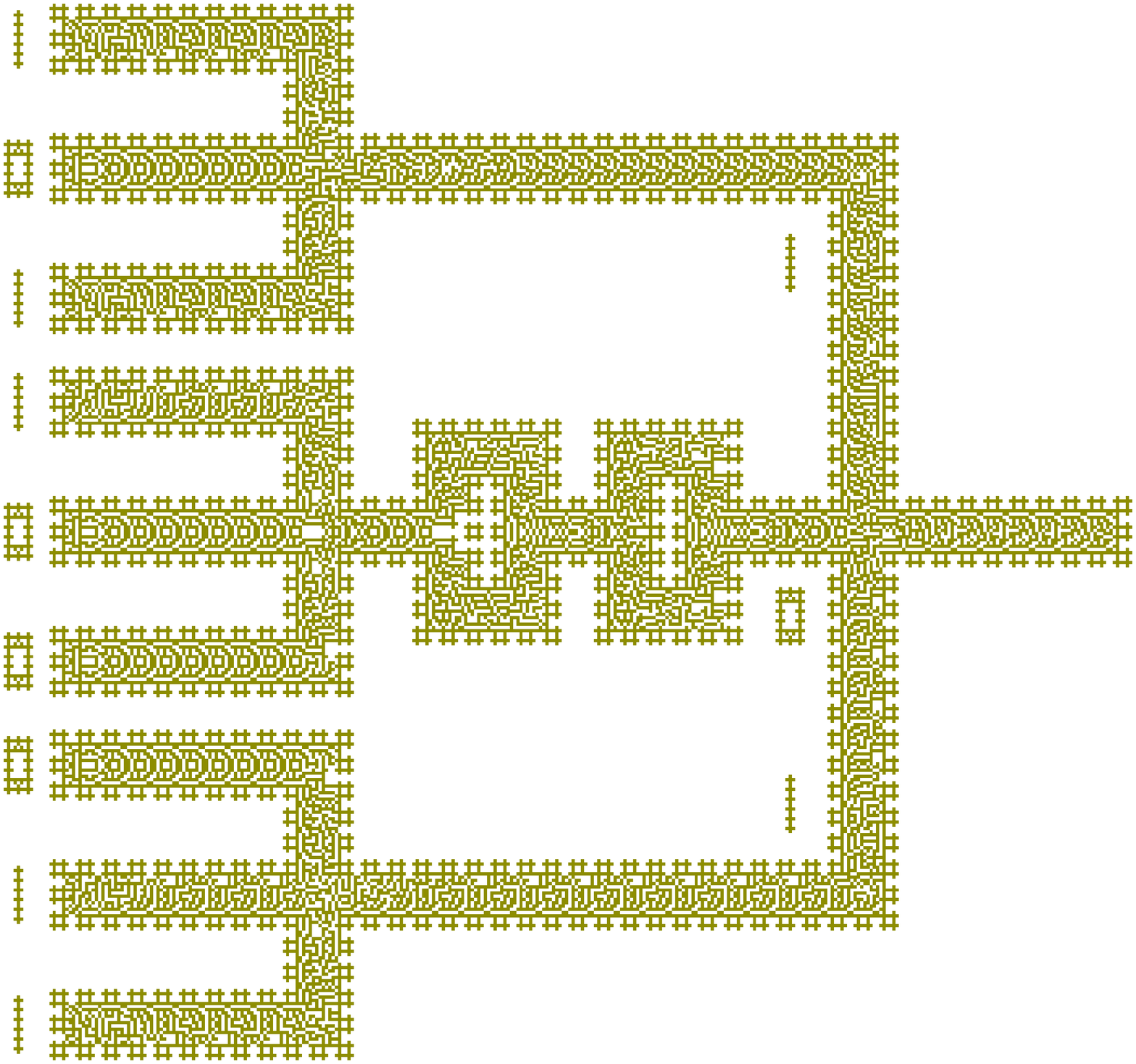}
\caption{Three {\sc majority} gate $MAJ(MAJ(1,0,1), MAJ(1,0,0), MAJ(0,1,$ $1)) \rightarrow 1$. The device is implemented in a space of $334 \times 326$ cells with 9,516 active cells, nine particles and two gates {\sc delay}. After 457 steps we obtain the final results with a population of 21,629 alive cells. All simulations where done with Golly \url{http://golly.sourceforge.net/}.}
\label{9WaveMajorityGate2}
\end{figure}

We can implement a cascade {\sc majority} gate by competing patterns initialized with three 3-input, shown in Fig. \ref{9WaveMajorityGate2}. It is composed by three 3-input {\sc majority} gates initialized with nine particles. In the process two gates {\sc delay} become necessary to synchronize the collision of patterns in the last 3-input {\sc majority} gate that give the final result. This tree {\sc majority} gates calculate specifically the inputs $MAJ(MAJ(1,0,1), MAJ(1,0,0), MAJ(0,1,1)) \rightarrow 1$. These types of gates are implemented in plasmonic devices~\cite{dutta2017proposal}.

\section{Final notes}

Computation by competing patterns is another unconventional way to design computers handling dozens, hundred, thousand or millions of organisms to interpret binary values in wires as fragments of wave propagation in a discrete space. A future work will be to implement other non-serial gates by competing patterns and develop other circuits with 5-input {\sc majority} gates and in cascade.

\bibliography{WaveMajorityGates_v01}

\begin{thebibliography}{10}

\bibitem{adamatzky2009hot}
Andrew Adamatzky.
\newblock Hot ice computer.
\newblock {\em Physics Letters A}, 374(2):264--271, 2009.

\bibitem{adamatzky2010physarum}
Andrew Adamatzky.
\newblock {\em Physarum machines: computers from slime mould}, volume~74.
\newblock World Scientific, 2010.

\bibitem{adamatzky2006phenomenology}
Andrew Adamatzky, Genaro~J. Mart{\'i}nez, and Juan~C. Seck-Tuoh-Mora.
\newblock Phenomenology of reaction--diffusion binary-state cellular automata.
\newblock {\em International Journal of Bifurcation and Chaos},
  16(10):2985--3005, 2006.

\bibitem{amlani1999digital}
Islamshah Amlani, Alexei~O. Orlov, Geza Toth, Gary~H. Bernstein, Craig~S. Lent,
  and Gregory~L. Snider.
\newblock Digital logic gate using quantum-dot cellular automata.
\newblock {\em Science}, 284(5412):289--291, 1999.

\bibitem{dieny2020opportunities}
Bernard Dieny, Ioan~Lucian Prejbeanu, Kevin Garello, Pietro Gambardella, Paulo
  Freitas, Ronald Lehndorff, Wolfgang Raberg, Ursula Ebels, Sergej~O
  Demokritov, Johan Akerman, et~al.
\newblock Opportunities and challenges for spintronics in the microelectronics
  industry.
\newblock {\em Nature Electronics}, 3(8):446--459, 2020.

\bibitem{dutta2017proposal}
Sourav Dutta, Odysseas Zografos, Surya Gurunarayanan, Iuliana Radu, Bart Soree,
  Francky Catthoor, and Azad Naeemi.
\newblock Proposal for nanoscale cascaded plasmonic majority gates for
  non-boolean computation.
\newblock {\em Scientific Reports}, 7(1):1--10, 2017.

\bibitem{eppstein2010growth}
David Eppstein.
\newblock Growth and decay in life-like cellular automata.
\newblock In {\em Game of Life Cellular Automata}, pages 71--97. Springer,
  2010.

\bibitem{fischer2017experimental}
Thomas Fischer, M~Kewenig, DA~Bozhko, AA~Serga, II~Syvorotka, Florin Ciubotaru,
  Christoph Adelmann, B~Hillebrands, and AV~Chumak.
\newblock Experimental prototype of a spin-wave majority gate.
\newblock {\em Applied Physics Letters}, 110(15):152401, 2017.

\bibitem{gao2017implementation}
Jinting Gao, Yaqing Liu, Xiaodong Lin, Jiankang Deng, Jinjin Yin, and Shuo
  Wang.
\newblock Implementation of cascade logic gates and majority logic gate on a
  simple and universal molecular platform.
\newblock {\em Scientific reports}, 7(1):1--7, 2017.

\bibitem{griffeath1996life}
David Griffeath and Cristopher Moore.
\newblock Life without death is p-complete.
\newblock {\em Complex Systems}, 10:437--448, 1996.

\bibitem{martinez2008logical}
Genaro~J. Mart{\'i}nez, Andrew Adamatzky, and Ben De~Lacy Costello.
\newblock On logical gates in precipitating medium: cellular automaton model.
\newblock {\em Physics Letters A}, 372(31):5115--5119, 2008.

\bibitem{martinez2010computation}
Genaro~J. Mart{\'i}nez, Andrew Adamatzky, Kenichi Morita, and Maurice
  Margenstern.
\newblock Computation with competing patterns in life-like automaton.
\newblock In {\em Game of Life Cellular Automata}, pages 547--572. Springer,
  2010.

\bibitem{martinez2010majority}
Genaro~J Mart{\'i}nez, Kenichi Morita, Andrew Adamatzky, and Maurice
  Margenstern.
\newblock Majority adder implementation by competing patterns in life-like rule
  b2/s2345.
\newblock In {\em Lecture Notes in Computer Science}, volume 6079, pages
  93--104. Springer, 2010.

\bibitem{mcintosh1990wolfram}
Harold~V McIntosh.
\newblock Wolfram's class iv automata and a good life.
\newblock {\em Physica D: Nonlinear Phenomena}, 45(1-3):105--121, 1990.

\bibitem{minsky1967computation}
Marvin Minsky.
\newblock Computation: Finite and infinite machines prentice hall.
\newblock {\em Inc., Engelwood Cliffs, NJ}, 1967.

\bibitem{navi2010five}
Keivan Navi, Samira Sayedsalehi, Razieh Farazkish, and Mostafa~Rahimi Azghadi.
\newblock Five-input majority gate, a new device for quantum-dot cellular
  automata.
\newblock {\em Journal of Computational and Theoretical Nanoscience},
  7(8):1546--1553, 2010.

\bibitem{ninagawa2021dynamics}
Shigeru Ninagawa.
\newblock Dynamics of self-reproducing cellular automata.
\newblock in preparation.

\bibitem{ninagawa1998flu}
Shigeru Ninagawa, Masaaki Yoneda, and Sadaki Hirose.
\newblock 1/f fluctuation in the “game of life”.
\newblock {\em Physica D: Nonlinear Phenomena}, 118(1-2):49--52, 1998.

\bibitem{prakash2019new}
G~Prakash, Mehdi Darbandi, N~Gafar, Noor~H Jabarullah, and Mohammad~Reza
  Jalali.
\newblock A new design of 2-bit universal shift register using rotated majority
  gate based on quantum-dot cellular automata technology.
\newblock {\em International Journal of Theoretical Physics}, 58(9):3006--3024,
  2019.

\bibitem{pudi2017majority}
Vikramkumar Pudi, K~Sridharan, and Fabrizio Lombardi.
\newblock Majority logic formulations for parallel adder designs at reduced
  delay and circuit complexity.
\newblock {\em IEEE transactions on computers}, 66(10):1824--1830, 2017.

\bibitem{radu2015spintronic}
IP~Radu, Odysseas Zografos, Adrien Vaysset, Florin Ciubotaru, Jingdong Yan,
  Johan Swerts, Dunja Radisic, Basoene Briggs, Bart Soree, Mauricio Manfrini,
  et~al.
\newblock Spintronic majority gates.
\newblock In {\em 2015 IEEE International Electron Devices Meeting (IEDM)},
  pages 32--5. IEEE, 2015.

\bibitem{rendell2016turing}
Paul Rendell.
\newblock {\em Turing machine universality of the Game of Life}.
\newblock Springer, 2016.

\bibitem{sasamal2016optimal}
Trailokya~Nath Sasamal, Ashutosh~Kumar Singh, and Anand Mohan.
\newblock An optimal design of full adder based on 5-input majority gate in
  coplanar quantum-dot cellular automata.
\newblock {\em Optik}, 127(20):8576--8591, 2016.

\bibitem{burks1966theory}
J.~von Neumann~(edited and completed~by A.~W.~Burks).
\newblock {\em Theory of self-reproducing automata}.
\newblock Urbana: University of Illinois Press, 1966.

\bibitem{wang2018novel}
Lei Wang and Guangjun Xie.
\newblock Novel designs of full adder in quantum-dot cellular automata
  technology.
\newblock {\em The Journal of Supercomputing}, 74(9):4798--4816, 2018.

\end{thebibliography}

\end{document}